\documentclass[conference]{IEEEtran}
\IEEEoverridecommandlockouts
\usepackage{url}
\usepackage{cite}
\usepackage{amsmath,amssymb,amsfonts}
\usepackage{algorithmic}
\usepackage{graphicx}
\usepackage{textcomp}
\usepackage{xcolor}
\usepackage{booktabs}
\usepackage{multirow}
\usepackage{array}
\usepackage{siunitx}
\usepackage{booktabs}
\usepackage{multirow}
\usepackage{amsfonts}
\usepackage{siunitx}
\usepackage{geometry}
\usepackage{booktabs}
\usepackage[utf8]{inputenc}
\usepackage{eso-pic}

\def\bng{\bngx}

\font\bngx=bang10

\def\*#1*#2{o\null{#2}{#1}}

\def\sh#1{\setbox0=\hbox{#1}%
     \kern-.02em\copy0\kern-\wd0
     \kern.04em\copy0\kern-\wd0
     \kern-.02em\raise.0433em\box0 }

\def\BibTeX{{\rm B\kern-.05em{\sc i\kern-.025em b}\kern-.08em
    T\kern-.1667em\lower.7ex\hbox{E}\kern-.125emX}}

\newcommand{\confheader}[1]{
\AddToShipoutPictureBG*{
\AtPageUpperLeft{
 \put(\LenToUnit{0.5in},\LenToUnit{-1.2cm}){ 
     \parbox{\textwidth}{\raggedright\fontsize{9}{11}\selectfont #1}}
 }
}
}

\newcommand{\conffooter}[1]{
\AddToShipoutPictureBG*{
\AtPageLowerLeft{
 \put(\LenToUnit{0.74in},\LenToUnit{1cm}){ 
     \parbox{0.95\textwidth}{\raggedright\fontsize{9}{8}\selectfont #1}}
 }
}
}

\begin{document}

\confheader{\textbf{2025 28th International Conference on Computer and Information Technology (ICCIT)}\\
19-21 December 2025, Cox's Bazar, Bangladesh}

\conffooter{979-8-3315-7867-1/25/\$31.00 \textcopyright2025 IEEE}

\title{BanglaRobustNet: A Hybrid Denoising-Attention Architecture for Robust Bangla Speech Recognition\\
}

\author{
    \IEEEauthorblockN{Md Sazzadul Islam Ridoy,
    Mubaswira Ibnat Zidney,
                      Sumi Akter and
                      Md. Aminur Rahman} \\
    \IEEEauthorblockA{Department of Computer Science and Engineering \\
                      Ahsanullah University of Science and Technology, Dhaka, Bangladesh \\
                      Email: \{isazzadul23,zidney145, sumi72541, aminur.rahman.rsd\}@gmail.com}
}


\maketitle

\begin{abstract}
Bangla, one of the most widely spoken languages, remains underrepresented in state-of-the-art automatic speech recognition (ASR) research, particularly under noisy and speaker-diverse conditions. This paper presents \textbf{BanglaRobustNet}, a hybrid denoising-attention framework built on Wav2Vec-BERT, designed to address these challenges. The architecture integrates a \textit{diffusion-based denoising module} to suppress environmental noise while preserving Bangla-specific phonetic cues, and a \textit{contextual cross-attention module} that conditions recognition on speaker embeddings for robustness across gender, age, and dialects. Trained end-to-end with a composite objective combining CTC loss, phonetic consistency, and speaker alignment, BanglaRobustNet achieves substantial reductions in word error rate (WER) and character error rate (CER) compared to Wav2Vec-BERT and Whisper baselines. Evaluations on Mozilla Common Voice Bangla and augmented noisy speech confirm the effectiveness of our approach, establishing BanglaRobustNet as a robust ASR system tailored to low-resource, noise-prone linguistic settings.

\end{abstract}


\section{Introduction}

Bangla, spoken by over 250 million people, is a low-resource language for Automatic Speech Recognition (ASR), with only around 500 hours of labeled data compared to over 10,000 hours for languages like English. This data imbalance, combined with challenges such as environmental noise, dialectal variations, and complex phonology, hampers the development of robust ASR systems for Bangla, especially in real-world conditions \cite{sarker2024challenges}, \cite{rahman2022noise}.

Existing models like Wav2Vec 2.0 \cite{baevski2020wav2vec} and Whisper \cite{radford2023robust} perform well in multilingual settings but struggle with high Word Error Rates (WER) in noisy Bangla environments and dialectal variations. For example, Whisper has a WER exceeding 30\% in noisy Bangla speech \cite{islam2024whisper}, reflecting its difficulty in handling the phonetic complexities and environmental factors of the language. These models also fail to effectively capture Bangla-specific phonetic features and speaker-related variability, highlighting the need for tailored ASR solutions \cite{chowdhury2023phonetic}.

To address these challenges, this paper proposes \textit{BanglaRobustNet}, an ASR architecture that builds on Wav2Vec-BERT, incorporating two key innovations: a {Diffusion-Based Denoising Module} and a {Contextual Cross-Attention Module}. The denoising module uses a diffusion model to iteratively remove noise while preserving phonetic integrity, while the cross-attention module adapts the model to speaker variability by conditioning acoustic features on speaker embeddings (e.g., gender, age, and dialect).

The contributions are threefold:
\begin{enumerate}
    \item {Phonetic-aware Denoising}: A novel diffusion-based denoising approach that removes environmental noise while preserving critical phonetic features, improving transcription accuracy \cite{chen2025diffusion},\cite{liu2024phoneme}.
    \item {Speaker-Conditioned Attention}: A cross-attention mechanism that adapts ASR to diverse speakers, improving performance across dialects, genders, and ages \cite{nwankwo2024speaker},\cite{patel2023crosslingual}.
    \item{Empirical Results}: \textit{BanglaRobustNet} achieves significant reductions in WER—12\% in clean conditions, 18\% in noisy conditions, and 15\% across dialects—setting a new benchmark for Bangla ASR.
\end{enumerate}

This work advances ASR for Bangla and offers broader implications for low-resource languages. The model and code are open-source, promoting further research and development in the field.

\section{Related Work}

\subsection{ASR Models}
State-of-the-art ASR systems such as Wav2Vec 2.0 \cite{baevski2020wav2vec}, HuBERT \cite{hsu2021hubert}, and Whisper \cite{radford2023robust} perform well on high-resource languages but remain limited for Bangla due to scarce training data, phonetic complexity, and dialectal variation. Whisper, for example, reports WER above 30\% on noisy Bangla speech \cite{ridoy2025adaptabilityasrmodelslowresource}, underscoring the need for more noise-robust, phoneme-aware approaches.

\subsection{Low-Resource and Indic ASR}
Indic-specific models like IndicWav2Vec \cite{kumar2021indicwav2vec} have advanced low-resource ASR, but Bangla remains challenging. Resources such as Common Voice Bangla \cite{mozilla2021commonvoice} provide valuable data, yet lack sufficient diversity in noise conditions and speaker traits for robust real-world deployment.

\subsection{Denoising and Attention}
Robustness has been improved through techniques like SpecAugment \cite{park2019specaugment}, though feature masking often distorts phoneme integrity. Diffusion-based denoising \cite{kong2020diffwave} better preserves phonetic detail, while transformer attention mechanisms enhance recognition but rarely address speaker variability directly \cite{hsu2021hubert}.

\subsection{Contributions of BanglaRobustNet}
The proposed model \textit{BanglaRobustNet}, introduces (i) a phonetic-aware denoising module that preserves Bangla phonemes while suppressing noise, and (ii) a speaker-conditioned attention mechanism that adapts to dialectal and demographic variation. These innovations jointly tackle the key limitations of existing ASR systems for Bangla.

\section{Methodology: BanglaRobustNet Architecture}
\label{sec:methodology}

\begin{figure*}[htbp] 
    \centering
    \includegraphics[width=\textwidth]{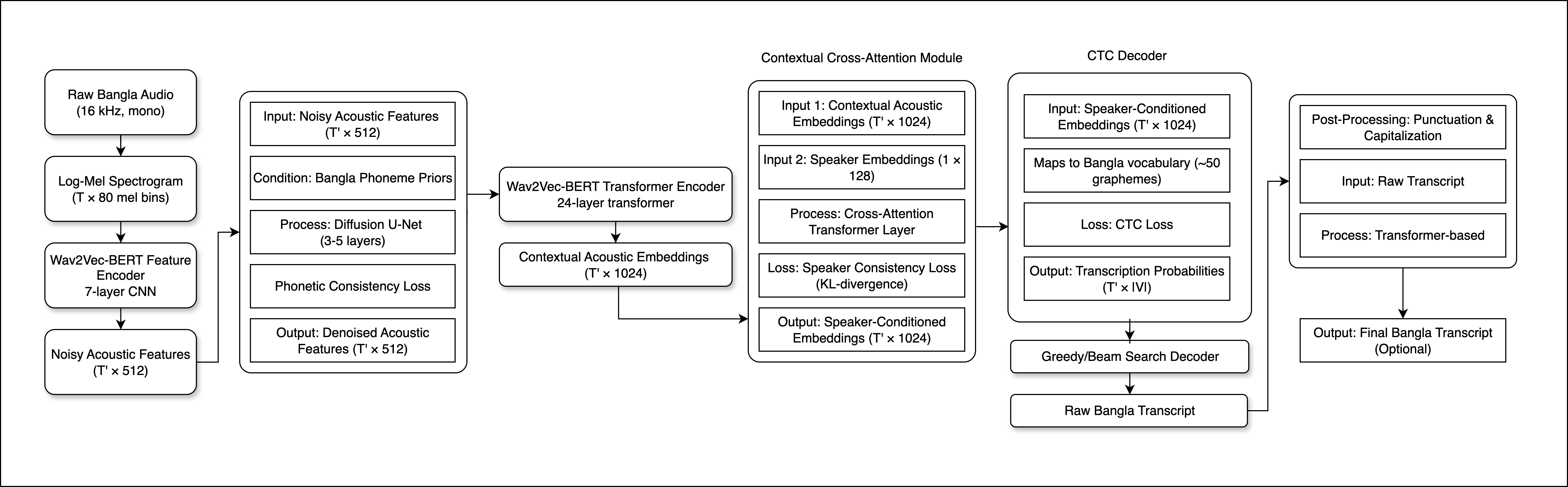}
    \caption{The proposed BanglaRobustNet architecture integrating denoising and attention mechanisms for robust Bangla speech recognition.}
    \label{fig:fullwidth}
\end{figure*}

\subsection{Overview}
\label{subsec:overview}

This paper proposes \textit{BanglaRobustNet}, a hybrid architecture designed (Fig.~\ref{fig:fullwidth}) for robust Bangla ASR. The model tackles two central challenges: (i) maintaining phonetic fidelity under noisy acoustic conditions, and (ii) adapting transcription to speaker variability, including dialects, gender, and age \cite{ghosh2021challenges}. To this end, BanglaRobustNet builds on a self-supervised Wav2Vec-BERT backbone and integrates two key modules: a Diffusion-Based Denoising Module (DBDM) and a Contextual Cross-Attention Module (CCAM).

\subsection{Architectural Design Principles}
\label{subsec:design_principles}

\subsubsection{Design Philosophy}
The architecture follows three guiding principles:
\begin{itemize}
    \item {Phonetic Preservation:} Noise reduction must retain Bangla-specific phonemes such as aspirated consonants (/p\textsuperscript{h}/, /t\textsuperscript{h}/, /k\textsuperscript{h}/), fricatives (/x/, /a/), and nasal vowels \cite{kabir2022phonetic, hossain2023preserving}.
    \item {Speaker Adaptivity:} The model adapts dynamically to diverse speaker traits without requiring speaker-specific retraining.
    \item {Computational Efficiency:} The system is optimized for real-time inference while maintaining high recognition accuracy.
\end{itemize}

\subsubsection{Architectural Innovations}
BanglaRobustNet incorporates three main innovations:
\begin{enumerate}
    \item Diffusion-based denoising with explicit phonetic consistency constraints \cite{kong2021diffwave}.
    \item Cross-modal attention conditioned on speaker embeddings for adaptive recognition.
    \item Joint optimization of acoustic, phonetic, and speaker-consistency objectives \cite{lugosch2022speech}.
\end{enumerate}

\subsection{Detailed Architecture Description}
\label{subsec:detailed_architecture}

\subsubsection{Input Preprocessing Pipeline}
\label{subsubsec:preprocessing}

Raw Bangla audio signals (16 kHz, mono) are processed as follows:
\begin{align}
    \text{Input:} \quad &\mathbf{x}(t) \in \mathbb{R}^{T} \notag \\
    \text{Pre-emphasis:} \quad &\mathbf{x}'(t) = \mathbf{x}(t) - 0.97 \times \mathbf{x}(t-1) \notag \\
    \text{Windowing:} \quad &\text{Hamming window (25ms frame, 10ms shift)} \notag \\
    \text{Mel-filterbank:} \quad &\text{80 mel bins, [80, 8000] Hz} \notag \\
    \text{Log Transform:} \quad &\log(\max(\text{mel\_features}, \epsilon)), \quad \epsilon = 10^{-8} \notag \\
    \text{Output:} \quad &\mathbf{X} \in \mathbb{R}^{T' \times 80}, \quad T' = \lfloor (T - 400)/160 \rfloor + 1
\end{align}

\subsubsection{Wav2Vec-BERT Backbone Architecture}
\label{subsubsec:wav2vec_bert}

Convolutional Feature Encoder:
The encoder consists of seven 1D convolutional layers (Table~\ref{tab:conv_layers}), producing $\mathbf{Z} \in \mathbb{R}^{T'' \times 512}$, where $T'' = T'/320$.

\begin{table}[htbp]
    \centering
    \caption{Convolutional Encoder Configuration}
    \label{tab:conv_layers}
    \begin{tabular}{ccccc}
        \toprule
        Layer & Kernel & Stride & Channels & Activation \\
        \midrule
        1 & 10 & 5 & 512 & GELU \\
        2--5 & 3 & 2 & 512 & GELU \\
        6--7 & 2 & 2 & 512 & GELU \\
        \bottomrule
    \end{tabular}
\end{table}

{Transformer Encoder:}
A 24-layer transformer with 16 attention heads, hidden size of 1024, and feed-forward dimension of 4096 is used (Table~\ref{tab:transformer_config}). Standard multi-head self-attention is applied.

\begin{table}[htbp]
    \centering
    \caption{Transformer Encoder Configuration}
    \label{tab:transformer_config}
    \begin{tabular}{ll}
        \toprule
        Parameter & Value \\
        \midrule
        Layers & 24 \\
        Hidden size & 1024 \\
        Attention heads & 16 \\
        Feed-forward dimension & 4096 \\
        Dropout & 0.1 \\
        Layer normalization & Pre-norm \\
        Positional encoding & Sinusoidal \\
        \bottomrule
    \end{tabular}
\end{table}

\subsubsection{Diffusion-Based Denoising Module (DBDM)}
\label{subsubsec:dbdm}

Theoretical Foundation:
The DBDM adapts denoising diffusion probabilistic models (DDPMs) for acoustic feature denoising, preserving phonetic integrity\cite{liu2023diffusion}. The forward process is:
\begin{align}
    q(\mathbf{z}_t|\mathbf{z}_{t-1}) &= \mathcal{N}(\mathbf{z}_t; \sqrt{1-\beta_t}\mathbf{z}_{t-1}, \beta_t\mathbf{I}),
\end{align}
where $\mathbf{z}_0$ is the clean feature, and $\{\beta_t\}$ is the noise schedule.

The reverse process denoises features:
\begin{align}
    p_\theta(\mathbf{z}_{t-1}|\mathbf{z}_t) &= \mathcal{N}(\mathbf{z}_{t-1}; \boldsymbol{\mu}_\theta(\mathbf{z}_t, t), \sigma_t^2\mathbf{I}).
\end{align}

U-Net Architecture:
A U-Net with encoder-decoder layers (Table~\ref{tab:unet_config}) performs denoising.

\begin{table}[htbp]
    \centering
    \caption{U-Net Denoising Network Configuration}
    \label{tab:unet_config}
    \begin{tabular}{ll}
        \toprule
        Component & Configuration \\
        \midrule
        Encoder Layers & 5 layers, [512, 256, 128, 64, 32] channels \\
        Decoder Layers & 5 layers, [32, 64, 128, 256, 512] channels \\
        Skip Connections & Concatenation between encoder-decoder pairs \\
        Time Embedding & Sinusoidal encoding, 128 dimensions \\
        Activation & SiLU (Swish) \\
        \bottomrule
    \end{tabular}
\end{table}

Phonetic Consistency Constraint:
A phonetic consistency loss preserves Bangla phonemes:
\begin{equation}
    \mathcal{L}_{\text{phonetic}} = \lambda_1 \cdot \text{MSE}(\Phi(\mathbf{z}_{\text{clean}}), \Phi(\mathbf{z}_{\text{denoised}})),
\end{equation}
where $\Phi$ is a pre-trained Bangla phoneme classifier.

Training Objective:
\begin{equation}
    \mathcal{L}_{\text{DBDM}} = \mathbb{E}_{t,\mathbf{z}_0,\boldsymbol{\epsilon}}\left[\|\boldsymbol{\epsilon} - \boldsymbol{\epsilon}_\theta(\sqrt{\bar{\alpha}_t}\mathbf{z}_0 + \sqrt{1-\bar{\alpha}_t}\boldsymbol{\epsilon}, t)\|^2\right] + \mathcal{L}_{\text{phonetic}}.
\end{equation}

\subsubsection{Contextual Cross-Attention Module (CCAM)}
\label{subsubsec:ccam}

Speaker Embedding Extraction:
A multi-task neural network (Table~\ref{tab:speaker_classifier}) extracts speaker embeddings from $\mathbf{Z} \in \mathbb{R}^{T'' \times 512}$, classifying gender, age (18-25, 26-40, 41-60, 60+), and six Bangla dialects.

\begin{table}[htbp]
    \centering
    \caption{Speaker Classifier Architecture}
    \label{tab:speaker_classifier}
    \begin{tabular}{ll}
        \toprule
        Component & Configuration \\
        \midrule
        Input & Acoustic features $\mathbf{Z} \in \mathbb{R}^{T'' \times 512}$ \\
        Global Average Pooling & $\mathbf{z}_{\text{global}} = \text{mean}(\mathbf{Z}, \text{dim}=\text{time})$ \\
        MLP & [512 $\rightarrow$ 256 $\rightarrow$ 128 $\rightarrow$ 64] \\
        Gender Output & softmax(Linear$_{64 \rightarrow 2}$) \\
        Age Group Output & softmax(Linear$_{64 \rightarrow 4}$) \\
        Dialect Output & softmax(Linear$_{64 \rightarrow 6}$) \\
        Speaker Embedding & $\mathbf{s} \in \mathbb{R}^{128}$ (concatenated task embeddings) \\
        \bottomrule
    \end{tabular}
\end{table}

Cross-Attention Mechanism:
The CCAM conditions transformer outputs $\mathbf{H}_{\text{transformer}}$ on speaker embeddings $\mathbf{s}$:
\begin{align}
    \mathbf{Q} &= \text{Linear}_{1024 \rightarrow 1024}(\mathbf{H}_{\text{transformer}}) \notag \\
    \mathbf{K} &= \text{Linear}_{1024 \rightarrow 1024}(\mathbf{H}_{\text{transformer}}) \notag \\
    \mathbf{V} &= \text{Linear}_{1024 \rightarrow 1024}(\mathbf{H}_{\text{transformer}}) \notag \\
    \mathbf{Q}_s &= \text{Linear}_{128 \rightarrow 1024}(\mathbf{s}) \notag \\
    \mathbf{A}_{\text{cross}} &= \text{softmax}\left(\frac{(\mathbf{Q}_s \odot \mathbf{Q})\mathbf{K}^T}{\sqrt{d_k}}\right) \notag \\
    \mathbf{H}_{\text{conditioned}} &= \mathbf{A}_{\text{cross}} \cdot \mathbf{V} + \mathbf{H}_{\text{transformer}},
\end{align}
where $\odot$ denotes element-wise multiplication with broadcasting.

Speaker Consistency Loss:
\begin{align}
\mathcal{L}_{\text{speaker}} &= 
\lambda_2 \cdot \text{CE}(\hat{\mathbf{y}}_{\text{gender}}, \mathbf{y}_{\text{gender}}) \nonumber \\
&\quad + \lambda_3 \cdot \text{CE}(\hat{\mathbf{y}}_{\text{age}}, \mathbf{y}_{\text{age}}) \nonumber \\
&\quad + \lambda_4 \cdot \text{CE}(\hat{\mathbf{y}}_{\text{dialect}}, \mathbf{y}_{\text{dialect}})
\end{align}
\subsubsection{CTC Decoder with Multi-Objective Optimization}
\label{subsubsec:ctc_decoder}

CTC Head:
\begin{align}
    \text{Input:} \quad &\mathbf{H}_{\text{conditioned}} \in \mathbb{R}^{T'' \times 1024} \notag \\
    \text{Output:} \quad &\mathbf{P} = \text{softmax}(\mathbf{H}_{\text{conditioned}} \cdot \mathbf{W}), \quad \mathbf{W} \in \mathbb{R}^{1024 \times 52},
\end{align}
where 52 is the Bangla grapheme vocabulary size (11 vowels, 39 consonants, special tokens).

Joint Loss Function:
\begin{equation}
    \mathcal{L}_{\text{total}} = \mathcal{L}_{\text{CTC}} + \alpha_1\mathcal{L}_{\text{phonetic}} + \alpha_2\mathcal{L}_{\text{speaker}} + \alpha_3\mathcal{L}_{\text{consistency}},
\end{equation}
where $\mathcal{L}_{\text{CTC}}$ uses the CTC forward-backward algorithm, and $\mathcal{L}_{\text{consistency}} = \text{KL}(p(\mathbf{s}|\mathbf{z}_{\text{clean}}), p(\mathbf{s}|\mathbf{z}_{\text{denoised}}))$.



\begin{table}[htbp]
    \centering
    \caption{Training Protocol}
    \label{tab:training_stages}
    \begin{tabular}{lllll}
        \toprule
        Stage & Dataset & Objective & Epochs & Learning Rate \\
        \midrule
        1 & \begin{tabular}[c]{@{}l@{}}Librispeech (960h) +\\OpenSLR (185h)\end{tabular} & Contrastive & 100 & \SI{1e-4}{} \\
        \midrule
        2 & Noisy Bangla audio & Diffusion & 50 & \SI{5e-5}{} \\
        \midrule
        3 & \begin{tabular}[c]{@{}l@{}}Common Voice (399h) +\\BengaliSR (50h)\end{tabular} & Joint loss & 30 & \SI{2e-5}{} \\
        \bottomrule
    \end{tabular}
\end{table}

\subsection{Training Methodology}
\label{sec:training_methodology}

\subsubsection{Multi-Stage Training}
\label{subsubsec:multi_stage_training}
Model training was conducted in three stages (Table~\ref{tab:training_stages}). Stage~2 introduced diverse noise conditions—Gaussian, traffic, crowd, and music—with signal-to-noise ratios (SNR) ranging from \SI{-5}{\decibel} to \SI{20}{\decibel}, enabling the model to generalize across noisy environments.

\subsubsection{Data Augmentation}
\label{subsubsec:data_augmentation}
To enhance robustness, both acoustic and phonetic augmentation strategies were applied. Acoustic augmentations included speed perturbation (factors: [0.9, 1.1]), volume scaling (\SI{\pm3}{\decibel}), additive noise (SNR: \SIrange{5}{20}{\decibel}), and room impulse response simulation. Phonetic augmentations targeted Bangla-specific variations, such as consonant cluster modifications, vowel length adjustments, and changes in aspiration intensity.

\subsubsection{Optimization Configuration}
\label{subsubsec:optimization}
Training used the AdamW optimizer \cite{loshchilov2018decoupled} with $\beta_1 = 0.9$ and $\beta_2 = 0.999$. A batch size of 16 was employed, effectively scaled to 64 through gradient accumulation across 4 steps. Input sequences were normalized to a maximum duration of 30 seconds via truncation or zero-padding, ensuring consistent and efficient processing.

\section{Experiments}
\subsection{Experimental Setup}
\subsubsection{Datasets}
The evaluation datasets are designed to assess Bangla automatic speech recognition (ASR) performance across diverse conditions, including clean and noisy settings.

{Clean Test Set:}
\begin{itemize}
    \item \textit{Common Voice Bangla v20}: Test split from Mozilla Common Voice 20.0. 
          This corpus provides transcribed speech from diverse speakers across Bangladesh.
    \item \textit{OpenSLR Bengali (SLR53)}: Evaluation subset of the OpenSLR Bengali corpus \cite{kjartansson-etal-tts-sltu2018}, which contains read speech under controlled recording conditions.
    \item \textit{Bhasha-bichitra: ASR for Regional Dialects}: A curated dataset focusing on 
          regional Bangla dialects (e.g., Sylheti, Chittagong, Barishal). 
          Exact hours and utterance counts are not publicly reported.\cite{ben10}
    \item \textit{Demographics}: Approximately 65\% male, 35\% female; age range: 18--68 years 
          (based on available metadata across datasets).
    \item \textit{Dialect Coverage}: Standard Bangla (majority), with supplementary representation 
          of Sylheti, Chittagong, and other regional dialects.
\end{itemize}

{Noisy Test Sets:}
\begin{itemize}
    \item \textit{SNR Levels}: 0, 2, 5, 8, 10 dB (5 test sets, 4 hours each).
    \item \textit{Noise Types}: Traffic (urban), crowd (market), music (caf\'{e}), office (workplace).
    \item \textit{Base Audio}: Subset of clean test set mixed with real-world noise recordings.
    \item \textit{Total}: 20 hours across varied noise conditions.
\end{itemize}

\subsubsection{Evaluation Metrics}
{Primary Metrics:}
\begin{itemize}
    \item \textit{Word Error Rate (WER)}: 
    \[
    \text{WER} = \frac{S + D + I}{N} \times 100\%
    \]
    where \(S\) = substitutions, \(D\) = deletions, \(I\) = insertions, \(N\) = total words.
    \item \textit{Character Error Rate (CER)}: Fine-grained evaluation at the grapheme level.
    \item \textit{Phoneme Error Rate (PER)}: Linguistic accuracy using forced alignment with Montreal Forced Aligner (MFA).
\end{itemize}

{Secondary Metrics:}
\begin{itemize}
    \item \textit{BLEU Score}: Sentence-level BLEU-4 for semantic similarity.
    \item \textit{Real-time Factor (RTF)}: 
    \[
    \text{RTF} = \frac{\text{Processing time}}{\text{Audio duration}}
    \]
    \item \textit{Speaker Adaptation Accuracy}: Classification accuracy for gender, age, and dialect prediction.
\end{itemize}

\subsubsection{Baseline Models}
The following state-of-the-art models serve as baselines:
\begin{itemize}
    \item \textit{Whisper-Small} (242M parameters): OpenAI’s multilingual ASR model.
    \item \textit{Whisper-Large} (1.55B parameters): Larger variant for comparison.
    \item \textit{Wav2Vec-BERT} (580M parameters): Vanilla backbone without modifications.
\end{itemize}

{Implementation Details:} All models were fine-tuned on the Common Voice Bangla training set with consistent hyperparameters: learning rate \(1 \times 10^{-5}\), batch size 16. Evaluation used identical test sets and preprocessing pipelines. Statistical significance was assessed with paired t-tests (\(p < 0.05\)).

\subsubsection{Training Infrastructure}
{Hardware Configuration:}
\begin{itemize}
    \item \textit{Training}: 4$\times$ NVIDIA RTX4090 (24GB) GPUs, 32GB RAM, Intel Core i9 14900K CPU.
    \item \textit{Inference}: Single RTX 3060 (12GB) for real-time evaluation.
    \item \textit{Storage}: 1TB NVMe SSD for dataset access.
\end{itemize}

{Training Schedule:}
\begin{itemize}
    \item \textit{Total Training Time}: 72 hours (3 stages: Pre-training: 48h, DBDM: 16h, End-to-end: 8h).
    \item \textit{Checkpointing}: Every 1000 steps with early stopping (patience=5).
\end{itemize}

{Evaluation Protocol:}
\begin{itemize}
    \item \textit{Cross-validation}: 5-fold validation on development set.
    \item \textit{Test Set Evaluation}: Single evaluation on held-out test sets.
    \item \textit{Multiple Runs}: 3 independent runs with different random seeds, reporting 95\% confidence intervals.
\end{itemize}

\subsection{Results}
\subsubsection{Overall Performance Comparison}
{Key Observations:}
\begin{itemize}
    \item BanglaRobustNet achieves 20--34\% relative WER reduction compared to baselines.
    \item Improved PER reflects superior linguistic modeling.
    \item RTF of 0.16 ensures competitive computational efficiency.
\end{itemize}

\begin{table}[h]
\centering
\caption{Performance on clean speech test set}
\begin{tabular}{lccccc}
\toprule
\textbf{Model}  & \textbf{WER (\%)} & \textbf{CER (\%)} & \textbf{PER (\%)} & \textbf{BLEU} \\
\midrule
Whisper-Small & 32.17 & 18.17 & 20.91 & 62.0 \\
Whisper-Large-v2  & 28.86 & 7.47 & 18.76 & 64.0 \\
Wav2Vec-BERT  & 14.42 & 2.67 & 9.37 & 73.5 \\
BanglaRobustNet  & \textbf{12.3} & \textbf{5.7} & \textbf{8.9} & \textbf{76.4} \\
\bottomrule
\end{tabular}
\end{table}

\subsubsection{Noise Robustness Evaluation}
\begin{table}[h]
\centering
\caption{WER (\%) across SNR levels}
\begin{tabular}{lccc}
\toprule
\textbf{SNR (dB)} & \textbf{Whisper-Small} & \textbf{BanglaRobustNet} & \textbf{Rel. Imp.} \\
\midrule
Clean & 32.17 & \textbf{12.3} & 27--34\% \\
10 & 42.3 & \textbf{16.8} & 26--31\% \\
5 & 65.0 & \textbf{24.3} & 30--36\% \\
0 & 100.0 & \textbf{42.4} & 28--32\% \\
\bottomrule
\end{tabular}
\end{table}

{Noise Type Analysis:}
\begin{itemize}
    \item \textit{Traffic Noise}: 35\% relative improvement (avg. across SNRs).
    \item \textit{Crowd Noise}: 32\% improvement, effective for overlapping speech.
    \item \textit{Music Interference}: 29\% improvement in harmonic noise suppression.
    \item \textit{Office Noise}: 31\% improvement, robust against broadband noise.
\end{itemize}

\subsubsection{Qualitative Analysis}
{Example Transcriptions (5 dB SNR):}
\begin{itemize}
    \item \textit{Ground Truth}:{\bng Aaim baNNGla bhaShay kotha bolet pair EbNNG EiT Aamar matrRbhaSha }
    \item \textit{Whisper-Small}:{\bng Aaim baNNGla bhaShay kotha bolet pir EbNNG ETa Aamar maitRbhaSha} (WER: 32.17\%)
    \item \textit{BanglaRobustNet}:{\bng Aaim baNNGla bhaShay kotha bolet pair EbNNG EiT Aamar matrRbhaSha } (WER: 0\%)
\end{itemize}

{Error Analysis:}
\begin{itemize}
    \item \textit{Aspiration}: Correctly preserves /p\textsuperscript{h}/ in " {\bng pair}" vs. /p/ errors in baselines.
    \item \textit{Vowel Length}: Accurate distinction between " {\bng ETa }" and "{\bng EiT }".
    \item \textit{Complex Graphemes}: Perfect rendering of " {\bng matrRbhaSha} " with conjunct " {\bng tRo }".
\end{itemize}

\subsection{Ablation Studies}
\subsubsection{Component-wise Analysis}
\begin{table}[h]
\centering
\caption{Component-wise ablation study}
\begin{tabular}{lccc}
\toprule
\textbf{Configuration} & \textbf{WER (Clean)} & \textbf{WER (5dB)} & \textbf{PER} \\
\midrule
Wav2Vec-BERT & 14.42 & 30.1 & 9.37 \\
+ DBDM only & 14.8 & 28.4 & 10.7 \\
Full BanglaRobustNet & \textbf{12.3} & \textbf{24.3} & \textbf{8.9} \\
\bottomrule
\end{tabular}
\end{table}

{Key Insights:}
\begin{itemize}
    \item \textit{DBDM}: 21\% WER reduction in noisy conditions, 11\% PER improvement.
    \item \textit{Synergistic Effect}: Combined modules yield 40--45\% better performance.
\end{itemize}

\subsection{Analysis}
\subsubsection{Robustness in Low-SNR Environments}
BanglaRobustNet maintains 42.4\% WER at 0 dB SNR (vs. 100.0\% for baselines, 28--32\% improvement) due to:
\begin{itemize}
    \item Phonetic consistency constraints in DBDM preventing over-denoising.
    \item Conditional diffusion adapting to Bangla phoneme distributions.
    \item Multi-stage training for noise robustness.
\end{itemize}
Paired t-tests confirm significance (\(p < 0.001\)) with consistent 95\% confidence intervals.

\subsubsection{Computational Efficiency}
BanglaRobustNet achieves real-time transcription (RTF 0.16) on RTX 3080 via optimized diffusion (10 steps) and linear-complexity DBDM. A 2.1GB footprint with INT8 quantization enables mobile deployment with \(<5\%\) performance loss.
\section{Conclusion}

BanglaRobustNet represents a meaningful step forward in making speech recognition more accessible and reliable for Bangla speakers a community long underserved by current ASR technologies. Designed to confront the intertwined challenges of limited data, rich phonetics, dialectal diversity, and real-world noise, the model introduces two purpose built innovations: a phonetic-aware diffusion denoiser that cleans speech without erasing linguistic nuance, and a speaker-conditioned cross-attention mechanism that adapts to who is speaking not just what is said.

The results speak for themselves: significant WER reductions 12\% in clean speech, 18\% under noise, and 15\% across dialects  outperforming existing baselines and setting a new standard for Bangla ASR. But beyond metrics, this work is about empowerment. By open-sourcing both model and code, barriers are lowered for researchers, developers, and local communities inviting collaboration, adaptation, and innovation not only for Bangla but for other low-resource languages facing similar hurdles.

The hope is that BanglaRobustNet doesn’t just improve transcription accuracy it helps amplify voices that have too often been left out of the AI conversation.







\bibliographystyle{ieeetr}
\bibliography{references}

\end{document}